\documentclass[
reprint,
superscriptaddress,
longbibliography,
bibnotes,
amsmath,amssymb,
aps,
prb,
]{revtex4-2}
\usepackage{graphicx}
\usepackage{bm}
\usepackage{dsfont}
\usepackage[usenames,dvipsnames]{xcolor}
\usepackage{pstricks}
\usepackage[tight]{subfigure}
\usepackage{verbatim}
\usepackage{units}
\usepackage{multirow}
\usepackage{enumitem}
\usepackage{mathrsfs}
\usepackage{leftidx}
\usepackage{xspace}
\usepackage{braket}
\usepackage{bbm}
\usepackage{cancel}
\usepackage{comment}
\usepackage{amsfonts,amssymb,amsmath}
\usepackage[normalem]{ulem}
\usepackage{hyperref}
\hypersetup{colorlinks=true,linktoc=all,linkcolor=blue,breaklinks=true,citecolor=blue,urlcolor=blue}
\usepackage[english]{babel}
\newcommand{\subfigref}[2]{\ref{#1}\hyperref[#1]{(#2)}}

\newcommand{\A}{\mathcal{A}}

\renewcommand{\S}{\mathcal{S}}
\newcommand{\T}{\mathcal{T}}
\newcommand{\U}{\mathcal{U}}
\newcommand{\W}{\mathcal{W}}

\newcommand{\Tr}[2]{\text{Tr}_{#1}\left\{#2\right\}}

\renewcommand{\L}{_\text{L}}
\newcommand{\R}{_\text{R}}
\newcommand{\Sys}{_\text{S}}
\newcommand{\B}{_\text{B}}

\begin{document}
\title{Thermodynamic and energetic constraints on transition probabilities of small-scale quantum systems}

\author{Ludovico Tesser}
\affiliation{Department of Microtechnology and Nanoscience (MC2), Chalmers University of Technology, S-412 96 G\"oteborg, Sweden}
\author{Matteo Acciai}
\affiliation{Scuola Internazionale Superiore di Studi Avanzati, Via Bonomea 265, 34136, Trieste, Italy}
\affiliation{Department of Microtechnology and Nanoscience (MC2), Chalmers University of Technology, S-412 96 G\"oteborg, Sweden}
 \author{Christian Sp\aa nsl\"att}
 \affiliation{Department of Engineering and Physics, Karlstad University, Karlstad, Sweden}
	\affiliation{Department of Microtechnology and Nanoscience (MC2), Chalmers University of Technology, S-412 96 G\"oteborg, Sweden}
 \author{In\`es Safi}
\affiliation{Laboratoire de Physique des Solides (UMR 5802),
CNRS-Universit\'e Paris-Sud and Paris-Saclay, Bâtiment 510, 91405 Orsay, France}

\author{Janine Splettstoesser}
	\affiliation{Department of Microtechnology and Nanoscience (MC2), Chalmers University of Technology, S-412 96 G\"oteborg, Sweden}
	
\date{\today}

\begin{abstract}
We study the transition probabilities of a two-point measurement on a quantum system, initially prepared in a thermal state. 
We find two independent constraints on the difference between transition probabilities when the system is prepared at different temperatures, which both turn out to be particularly restrictive when the measured quantum system is small. 
These bounds take the form of a \textit{thermodynamic} and of an \textit{energetic} constraint, as they are associated with the dissipated heat and with the absorbed energy required to increase or to reduce the temperature of the system. The derived constraints apply to arbitrary system Hamiltonians, including interactions or non-linear energy spectra. 
We show the relevance of these constraints for the special case where transitions are induced by energy or particle exchange in weakly coupled bipartite systems out of equilibrium. This example is of interest for a wide range of experimentally relevant systems, from molecular junctions to coupled cavities, and can be tested by, for instance, measuring the out-of-equilibrium tunneling current and its noise.
\end{abstract}

\maketitle

\section{Introduction}\label{sec:introduction}

Fluctuation theorems have been instrumental in studying the probability distribution of physical variables, such as thermodynamic work, in both classical and quantum stochastic thermodynamics~\cite{Jarzynski1997Apr, Crooks1999Sep, Tasaki2000Sep, Andrieux2007Feb, Jarzynski2007Jun, Andrieux2009Apr, Esposito2009Dec, Sagawa2012Feb, Saira2012Oct, Bochkov2013Jun, Barato2015Apr, Potts2018Nov, Timpanaro2019Aug,Pekola2021Oct, Strasberg2022Book, Guarnieri2024Aug}. 
In particular, detailed fluctuation theorems~\cite{Esposito2009Dec,Landi2024Apr} constrain such probability distributions by relating the probability of a process to the probability of its time reverse.
These relations provide a powerful framework to study stochastic dynamics out of equilibrium, but they can also be used to, e.g., derive the equilibrium fluctuation-dissipation theorem (FDT)~\cite{Callen1951Jul, Green1954Mar, Kubo1957Jun}, which relates the fluctuations of observables to their dissipative responses.  
However, establishing a relation analogous to the FDT, linking generic correlations to response functions for systems \textit{out of equilibrium} remains challenging~\cite{Altaner2016Oct, Dechant2020Mar, Shiraishi2022Jul}.
For out-of-equilibrium correlated states, FDTs have been identified~\cite{safi2011Nov,Safi2014Jan,Safi2020Jul} for a generalized current operator, whose average and fluctuations are determined by two independent nonequilibrium transfer rates.
Specifically, for a charge current induced by a voltage bias, a FDT~\cite{Rogovin1974Jul,Levitov2004Sep} has been established far from equilibrium in the \emph{(weak) tunneling regime}. 
In essence, this FDT extension relies on the detailed balance relation between these rates under the crucial assumption of a uniform temperature across the tunneling link. Consequently, the generalized FDT breaks down in the presence of a more generic out-of-equilibrium situation, such as in the presence of a temperature bias. 

However, setups that can subsequently be in contact with environments at different temperatures, or even subject to a temperature bias are crucial in (quantum) thermodynamics, where they are used to fuel, e.g., heat engines~\cite{Benenti2017Jun,Whitney2019Apr,Cangemi2024Oct}. 
Pivotal experiments have not only implemented nanoscale heat engines~\cite{Thierschmann2015Oct,Josefsson2018Oct,Prete2019May,Guthrie2022Jun} but also explored temperature biases for transport spectroscopy~ \cite{safi_pierre_T_bias_2021,Gehring2021Apr,Esat2023Apr, Gemma2023Jun}. Importantly, systems exposed to large temperature biases also occur when one subsystem is cooled down with the help of a coupling to another, possibly very different, subsystem~\cite{Giazotto2006Mar,Aspelmeyer2014Dec,Senior2020Feb}.
It is hence important to understand in which way
 coupling a system to different temperatures constrains its dynamics.

\begin{figure}
    \centering
    \includegraphics[]{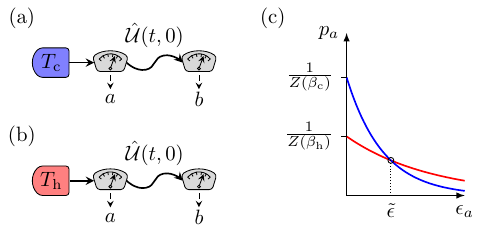}
    \caption{Two-point measurement schemes starting with an initially cold (a) or hot (b) state. The Gibbs probabilities \eqref{eq:gibbs-probability} of observing the eigenstate of energy $\epsilon_a$ for a cold (blue) or hot (red) initial state (c) only cross at one energy, which we refer to as $\tilde{\epsilon}$.}
    \label{fig:boltzmann-crossing}
\end{figure}

In this work, we present general relations between transition probabilities in a two-point measurement scheme, comparing situations where the system is initially prepared in thermal states at different temperatures, see Fig.~\ref{fig:boltzmann-crossing}. 
This is a different approach than those in earlier studies on fluctuation relations~\cite{Esposito2009Dec,Strasberg2022Book,Landi2024Apr}, where transitions of a process are compared to transitions of the time-reversed process. 
Our approach allows us to establish bounds on the difference between transition probabilities:
the main results of this paper are the thermodynamic constraint (Sec.~\ref{subsec:thermo-constraints}) and the energetic constraint (Sec.~\ref{subsec:energy-constraints}) on the temperature-dependent transition probabilities.  They relate the thermodynamic and energetic cost required to bring the system to these different temperatures with the response of the transition probabilities to a temperature variation. 
We do not make any assumptions on the Hamiltonian underlying the transition probabilities, which can hence involve strong interactions or nonlinear spectral properties.

One relevant application of our general findings are constraints on the dynamics of a bipartite system subject to a temperature bias. 
This could, e.g., be a small quantum system prepared at an initial temperature and coupled to an environment at a different temperature. As examples, we will choose bipartite systems with a weak tunnel coupling. The reason for this choice is the possibility to clearly identify the two different subsystem states and that the transition rates can be directly related to tunneling currents and the zero-frequency current noise~\cite{Rogovin1974Jul,Levitov2004Sep}.
This allows us to exploit our findings to formulate constraints on noise in the spirit of FDTs but in the presence of a possibly large temperature bias, where standard FDTs are not applicable~\cite{Callen1951Jul, Green1954Mar, Kubo1957Jun,Rogovin1974Jul,Levitov2004Sep}.

Indeed, in particular for nanoelectronic systems, fluctuation-dissipation bounds have recently been developed~\cite{Tesser2024May} for current fluctuations (or noise) in the presence of a temperature bias. Although these bounds apply to conductors with generic transmission properties, their validity is limited to systems with weak electron-electron interactions.
In the presence of possibly strong interactions, the perturbative approach developed in Refs.~\cite{Safi2014Jan,Safi2020Jul} showed that noise is super-Poissonian in the tunneling regime, even in the presence of a temperature bias. However, this constraint does not single out the
role of the temperature bias.
Hence, there remains the important question of whether fundamental bounds on the dynamics for strongly interacting systems exist accounting for the impact of a temperature bias.

For such systems, using the derived general thermodynamic and energetic constraints, we establish bounds on the nonequilibrium tunneling rates in the presence of a possibly large temperature bias, accounting for the thermodynamic quantities required to generate such bias.
These findings have direct implications on how the noise in temperature-biased systems is constrained by the system dynamics. We thereby extend the scope of out-of-equilibrium noise at the intersection of quantum transport and quantum thermodynamics to systems with possibly strong interactions.
Importantly, our findings do not rely on any close-to-equilibrium fluctuation theorems.

The remainder of this paper is organized as follows. In Sec.~\ref{sec:transition-probabilities}, we briefly introduce the concept of the two-point measurement with initial thermal states to then subsequently present a thermodynamic and an energetic constraints on the transition probabilities. In Sec.~\ref{sec:bipartite}, we apply our general results to experimentally relevant examples, which can be classified as weakly coupled bipartite systems in the presence of a large temperature bias. In Secs.~\ref{subsec:Jaynes-Cummings} and \ref{subsec:fermionic-rings}, we showcase the constraints for an atom coupled to a cavity and two coupled fermionic tight-binding rings. Several Appendices provide details of our derivations of key equations.

\section{Constraints on dynamics with different initial states}\label{sec:transition-probabilities}

Our goal is to compare the dynamics of two-point measurement schemes with different initial thermal states of the system and to constrain this difference by thermodynamic quantities. In this Section, we present general \textit{thermodynamic} and \textit{energetic} constraints.

\subsection{Transition probabilities}\label{subsec:transition-probabilities}

We consider the general case in which a quantum system, initially prepared in the state described by the density matrix $\hat{\rho}$, undergoes a two-point projective measurement process.
The first measurement is done on the basis $\{\ket{i}\}_i$, and has outcome $a$ with probability 
\begin{equation}\label{eq:initial_prob}
    p_a = \braket{a|\hat{\rho}|a} .
\end{equation}
Then, the system undergoes an arbitrary unitary evolution $\hat{\mathcal{U}}(t,0)$ until time $t$, when a second measurement takes place. This last measurement is done on the basis $\{\ket{\psi_i}\}_i$, which may differ from the one of the initial measurement.
The joint probability of measuring outcome $b$ in the second measurement, after the first measurement had outcome $a$, is given by
\begin{equation}\label{eq:transition_probability}
    p_{a\to b} = \left|\braket{\psi_b|\hat{\mathcal{U}}(t,0)|a }\right|^2 p_a,
\end{equation}
which is the probability of observing a transition $a\to b$ in the measurement outcomes.

These transition probabilities, compared with the transition probabilities in the time-reversed process are typically the starting point to develop fluctuation theorems~\cite{Esposito2009Dec,Strasberg2022Book,Landi2024Apr}. 
Here, we are interested in finding out the impact of \textit{different temperatures} on the dynamics of a system. We therefore start with the important statement that the temperatures only influence the initial state of the two-point measurement and not the conditional probabilities $|\braket{\psi_b|\hat{\mathcal{U}}(t,0)|a }|^2$. This means that we---instead of what is done in typical derivations of fluctuation theorems---need to compare the transition probabilities in the \textit{same process} (i.e., induced by the same unitary evolution), but with \textit{different initial states}.
In particular, we consider initial states $\hat{\rho}$ being Gibbs states,
\begin{equation}\label{eq:gibbs}
    \hat{\rho} = \frac{e^{-\beta \hat{H}_0}}{Z(\beta)}\,,
\end{equation}
where $\hat{H}_0$ is the system's Hamiltonian at time $t=0$, $\beta\equiv T^{-1}$ is the inverse temperature of the system (note that temperatures have the units of energy, meaning that we set $k_\text{B}=1$), and $Z(\beta)\equiv \Tr{}{e^{-\beta\hat{H}_0}}$ is the partition function. 
Furthermore, we take the first measurement to be in the energy eigenbasis $\hat{H}_0 \ket{a} = \epsilon_a\ket{a}$, such that the probability of observing outcome $a$ in Eq.~\eqref{eq:initial_prob} reads
\begin{equation}\label{eq:gibbs-probability}
    p_a(\beta) = \frac{e^{-\beta \epsilon_a}}{Z(\beta)}
\end{equation}
where $\epsilon_a$ is the energy of $\ket{a}$.
We highlight the dependence on the inverse temperature $\beta$ by putting it as an argument of $p_a$ because, in the following, we compare initial states at \textit{different temperatures}, as sketched in Fig.~\subfigref{fig:boltzmann-crossing}{a,b}. Using properties of the Gibbs distribution, we establish constraints on the transition probabilities of Eq.~\eqref{eq:transition_probability} in the sections below that contain thermodynamic quantities, such as the internal energy $U$ and the entropy $S$. Those are given by
\begin{subequations}
\begin{align}
    U(\beta)& \equiv \Tr{}{\hat{H}_0 \frac{e^{-\beta \hat{H}_0}}{Z(\beta)}},\label{eq:internal_U}\\
    S(\beta)&\equiv -\Tr{}{\frac{e^{-\beta \hat{H}_0}}{Z(\beta)}\ln \frac{e^{-\beta \hat{H}_0}}{Z(\beta)}} .\label{eq:entropy_sigma}
\end{align}
\end{subequations}
These quantities of the initial state are of interest for the constraints to be developed, since the temperature-dependence of the transition probabilities, Eq.~\eqref{eq:transition_probability}, enters via the initial-state probability, Eq.~\eqref{eq:initial_prob},  only.

\subsection{Thermodynamic constraint}\label{subsec:thermo-constraints}

\begin{figure}
    \centering
    \includegraphics{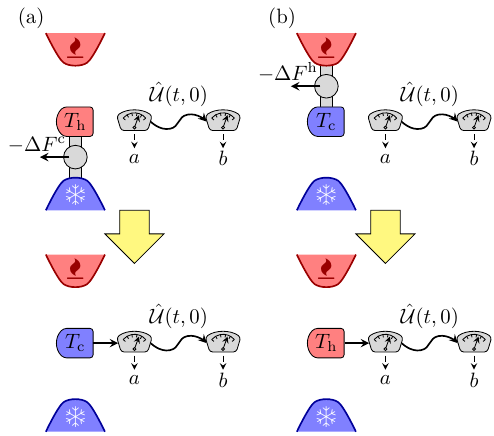}
    \caption{(a) An initially hot system is cooled down to $T_\text{c}$ using a cold bath. In this process, one can extract at most $-\Delta F^\text{c}$ [Eq.~\eqref{eq:DeltaFc}] as work. The now cold system undergoes the two-point measurement scheme. (b) An initially cold system is heated up to $T_\text{h}$ using a hot bath. In this process, one can extract at most $-\Delta F^\text{h}$ [Eq.~\eqref{eq:DeltaFh}] as work. The now hot system undergoes the two-point measurement scheme.}
    \label{fig:thermo-heat-cool}
\end{figure}

We start with an analysis of the initial state at different temperatures. When comparing a hot state at inverse temperature $\beta_\text{h}$ with a cold state at inverse temperature $\beta_\text{c}>\beta_\text{h}$, there is only one value of $\epsilon_a$ for which $p_a(\beta_\text{h})=p_a(\beta_\text{c})$, as shown in Fig.~\subfigref{fig:boltzmann-crossing}{c}.
This crossing energy is given by 
\begin{equation}
    \tilde{\epsilon} = \frac{1}{\beta_\text{c}- \beta_\text{h}}\ln\frac{Z(\beta_\text{h})}{Z(\beta_\text{c})}.
\end{equation}
Therefore the state-probabilities $p_a$ fulfill $p_a(\beta_\text{h})\gtrless p_a(\beta_\text{c})$ when $\epsilon_a\gtrless\tilde{\epsilon}$.
It follows that
\begin{equation}\label{eq:crossing-inequality}
    [p_a(\beta_\text{h}) - p_a(\beta_\text{c})][\epsilon_a - \tilde{\epsilon}] \geq 0
\end{equation}
for all $\epsilon_a$ and $\beta_\text{c}>\beta_\text{h}$.
Summing \eqref{eq:crossing-inequality} over all eigenstates $a$ leads to the statement that the internal energy is an increasing function of temperature, $U(\beta_\text{h})\geq U(\beta_\text{c})$. We furthermore note that the contributions of~\eqref{eq:crossing-inequality} can partially be rewritten in terms of a response of the probabilities to a temperature variation
\begin{equation}\label{eq:probability_response}
    \partial_\beta p_{a}(\beta) = [U(\beta)-\epsilon_a] p_{a}(\beta).
\end{equation}
In contrast, the remaining term $[U(\beta_\mathrm{h})- \tilde{\epsilon}]p_a(\beta_\text{h})-[U(\beta_\mathrm{c})- \tilde{\epsilon}] p_a(\beta_\text{c})$ then includes  thermodynamic quantities, in particular the nonequilibrium free energy differences,
defined as
\begin{subequations}
    \begin{align}
        \Delta F^\text{c} &\equiv [U(\beta_\text{c}) - U(\beta_\text{h})] -T_\text{c}[S(\beta_\text{c}) - S(\beta_\text{h})], \label{eq:DeltaFc}\\
        \Delta F^\text{h} &\equiv [U(\beta_\text{h}) - U(\beta_\text{c})] -T_\text{h}[S(\beta_\text{h}) - S(\beta_\text{c})].  \label{eq:DeltaFh}
    \end{align}
\end{subequations}
We find the relation
\begin{eqnarray}
&&[U(\beta_\mathrm{h})- \tilde{\epsilon}]p_a(\beta_\text{h})-[U(\beta_\mathrm{c})- \tilde{\epsilon}] p_a(\beta_\text{c})\nonumber\\
& = &   -\frac{\beta_\text{c}\Delta F^\text{c}}{\beta_\text{c}-\beta_\text{h}} p_{a}(\beta_\text{h}) - \frac{\beta_\text{h}\Delta F^\text{h}}{\beta_\text{c}-\beta_\text{h}}p_{a}(\beta_\text{c}).\label{eq:thermo-probability}
\end{eqnarray}
In this paper, we are interested in using the inequality \eqref{eq:crossing-inequality} to formulate constraints for \textit{transition} probabilities $p_{a\to b}$ in a two-point measurement, see Eq.~\eqref{eq:transition_probability}. The above discussed insights can be readily transferred to transition probabilities, by simply multiplying \eqref{eq:crossing-inequality}, as well as \eqref{eq:probability_response} and \eqref{eq:thermo-probability}, by the conditional probability $|\braket{\psi_b|\mathcal{U}(t,0)|a}|^2$, finding
\begin{equation}\label{eq:thermo-constraint-0}
    [p_{a\to b}(\beta_\text{h}) -p_{a\to b}(\beta_\text{c})][\epsilon_a - \tilde{\epsilon}]\geq 0.
\end{equation}
Using the properties of \eqref{eq:probability_response} and \eqref{eq:thermo-probability} in \eqref{eq:thermo-constraint-0}, we  establish the bound on transition probabilities
\begin{equation}\label{eq:thermo-constraint}
    \mathcal{W}_{a\to b}^\text{Thermo} \geq \mathcal{W}_{a\to b}^\text{Resp}\,,
\end{equation}
where we defined a temperature-response function
\begin{equation}\label{eq:response-term}
    \mathcal{W}_{a\to b}^\text{Resp} \equiv \partial_\beta p_{a\to b}(\beta_\text{h}) - \partial_\beta p_{a\to b}(\beta_\text{c}),
\end{equation}
and a thermodynamic cost function
\begin{equation}\label{eq:thermo-term}
    \mathcal{W}_{a\to b}^\text{Thermo} \equiv -\frac{\beta_\text{c}\Delta F^\text{c}}{\beta_\text{c}-\beta_\text{h}} p_{a\to b}(\beta_\text{h}) - \frac{\beta_\text{h}\Delta F^\text{h}}{\beta_\text{c}-\beta_\text{h}}p_{a\to b}(\beta_\text{c}).
\end{equation}
This \textit{thermodynamic constraint}~\eqref{eq:thermo-constraint} on the transition probabilities at different temperatures is the first main result of this paper. It implies that the response of the transition probabilities to a change in temperature is limited by the transition probabilities themselves and by the thermodynamic cost of changing the system's temperature. In order to intuitively understand how far it constrains the system dynamics at different initial temperatures, we consider the thermodynamics of cooling down a hot system in order to be able to subsequently perform the two-point measurement scheme starting from a cold state, as sketched in Fig.~\subfigref{fig:thermo-heat-cool}{a}. Analogously, we consider the thermodynamics of heating up a cold system in order to be able to subsequently perform the two-point measurement scheme starting from a hot state, as sketched in Fig.~\subfigref{fig:thermo-heat-cool}{b}. 
\begin{itemize}
    \item[(a)] \textit{Cooling:} The system initially at $T_\text{h}$ is cooled down by bringing it into contact with a bath at $T_\text{c}<T_\text{h}$ [Fig.~\subfigref{fig:thermo-heat-cool}{a}]. Heat flows out of the system until it reaches temperature $T_\text{c}$, thus inducing the change $[U(\beta_\text{c}) - U(\beta_\text{h})]$ in internal energy and $[S(\beta_\text{c}) - S(\beta_\text{h})]$ in entropy. While heat is flowing, it is possible to extract work, which is at most $-\Delta F^\text{c}$.
    After the system has been cooled down, the two-point measurement scheme is performed and leads to the transition probabilities $p_{a\to b}(\beta_\text{c})$.
    \item[(b)] \textit{Heating:} The system initially at $T_\text{c}$ is heated up by bringing it into contact with a bath at $T_\text{h}>T_\text{c}$ [Fig.~\subfigref{fig:thermo-heat-cool}{b}]. Heat flows into the system until it reaches temperature $T_\text{h}$, thus inducing the change $[U(\beta_\text{h}) - U(\beta_\text{c})]$ in internal energy and $[S(\beta_\text{h}) - S(\beta_\text{c})]$ in entropy. While heat is flowing, it is possible to extract work, which is at most $-\Delta F^\text{h}$.
    After the system has been heated up, the two-point measurement scheme is performed and leads to the transition probabilities $p_{a\to b}(\beta_\text{h})$.
\end{itemize}
Furthermore, the factors $\eta^{(\mathrm{h})}\equiv\frac{\beta_\text{c}}{\beta_\text{c}-\beta_\text{h}} = \frac{T_\text{h}}{T_\text{h}-T_\text{c}}$ and $\eta^{(\mathrm{c})}\equiv\frac{\beta_\text{h}}{\beta_\text{c}-\beta_\text{h}} = \frac{T_\text{c}}{T_\text{h}-T_\text{c}}$ in Eq.~\eqref{eq:thermo-term} correspond to the coefficient of performance of a heat pump and of a refrigerator, respectively.
Consequently, the product $-\frac{\beta_\text{c}\Delta F^{\text{c}}}{\beta_\text{c}-\beta_\text{h}} $ sets a lower limit on the heat absorbed by the cold bath during the cooling process when the extracted work is maximum, as detailed in Appendix~\ref{app:extracting-work}.
Similarly, the product $-\frac{\beta_\text{h}\Delta F^{\text{h}}}{\beta_\text{c}-\beta_\text{h}}$ sets a lower limit on the energy absorbed by the system during the heating process.
Thus, in the thermodynamic cost function of Eq.~\eqref{eq:thermo-term} the hot and cold transition probabilities are weighted by the heat dissipated to cool down or heat up the system, respectively.

\subsection{Energetic constraint}\label{subsec:energy-constraints}

\begin{figure}
    \centering
    \includegraphics{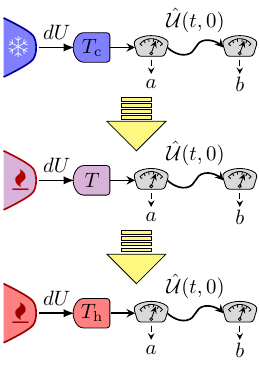}
    \caption{Heating stroke of the system. As the temperature increases from $T_\mathrm{c}$ to $T_\mathrm{h}$ (top to bottom panels), the energy absorbed, $dU$, and the transition probabilities, $p_{a\to b}$,  are monitored at each intermediate temperature.}
    \label{fig:energy-stroke}
\end{figure}
In order to set up the second constraint, instead of comparing the initial probability distribution, $p_a(\beta)$, at two different temperatures, we analyze how the distribution changes under an infinitesimal temperature variation. Differentiating Eq.~\eqref{eq:probability_response}, we have
\begin{equation}\label{eq:start-energetic}
    \partial_\beta^2 p_a(\beta) = p_a(\beta)[U(\beta) - \epsilon_a]^2  + p_a(\beta)\partial_\beta U(\beta).
\end{equation}
Since the first term on the right hand side is always positive, we can establish the inequality
\begin{equation}\label{eq:integral_energy_prob}
    \int_{U(\beta_\text{c})}^{U(\beta_\text{h})} p_a(\beta) [dU(\beta)] \geq \partial_\beta p_a (\beta_\text{h}) - \partial_\beta p_a(\beta_\text{c})\ .
\end{equation}
Multiplying Eq.~\eqref{eq:integral_energy_prob} by the conditional probability $|\braket{\psi_b|\mathcal{U}(t,0)|a}|^2$ reveals a constraint on the transition probability $p_{a\to b}$ of Eq.~\eqref{eq:transition_probability}, given by
\begin{equation}\label{eq:en-constraint}
 \mathcal{W}_{a\to b}^\text{Energy}\geq     \mathcal{W}_{a\to b}^\text{Resp}
\end{equation}
with the temperature-response function of Eq.~\eqref{eq:response-term} and the energetic cost function defined as 
\begin{equation}\label{eq:en-term}
     \mathcal{W}_{a\to b}^\text{Energy}\equiv  \int_{U(\beta_\text{c})}^{U(\beta_\text{h})} p_{a\to b}(\beta) [dU(\beta)].
\end{equation}
Equation~\eqref{eq:en-constraint} is the second central result of this paper. We refer to it as the \textit{energetic constraint} since---unlike the thermodynamic constraint of Eq.~\eqref{eq:thermo-constraint}---it does not focus on the thermodynamic cost required to generate the temperature bias, but rather on the energetic cost.
To understand the ingredients of this second constraint \eqref{eq:en-constraint}, one does not need to consider both heating and cooling processes depicted in Fig.~\ref{fig:thermo-heat-cool}, but only one continuous heating stroke as shown in Fig.~\ref{fig:energy-stroke}.
Starting from the system being at a cold temperature $T_\mathrm{c}=\beta_\mathrm{c}^{-1}$, we imagine slowly increasing its temperature to the hot temperature $T_\mathrm{h}=\beta_\mathrm{h}^{-1}$ and identifying the transition probabilities at each intermediate infinitesimal temperature change, as indicated by the interrupted yellow arrows. The energetic cost is found by weighting the transition probabilities with the corresponding variation of the internal energy needed to increase the temperature of the system.

\subsection{Saturating constraints and trivial constraints}\label{sec:equality}

The inequalities set up in Eq.~\eqref{eq:thermo-constraint} and Eq.~\eqref{eq:en-constraint} generically constrain the difference in transition probabilities in two-point measurement schemes for initial states at different temperatures. Here, we elaborate on the conditions under which these constraints are saturated and those under which they are trivial.
By construction, both thermodynamic and energetic bounds become equalities at equal temperature, $\beta_\mathrm{c}=\beta_\mathrm{h}$. However, in this regime Eqs.~(\ref{eq:thermo-constraint},~\ref{eq:en-constraint}) become trivial since $\W^\text{(Thermo)}_{a\to b}=\W^\text{(Energy)}_{a\to b}=\W^\text{(Resp)}_{a\to b}=0$. 

However, the constraints can be saturated non-trivially even for different temperatures when the energy of the initial state of the two-point measurement takes on specific values. The difference between Eq.~\eqref{eq:thermo-constraint} and Eq.~\eqref{eq:en-constraint} is also reflected in the conditions required to saturate the two constraints. For the thermodynamic constraint this energy corresponds to the energy at which the Gibbs distribution at different temperatures cross $p_a(\beta_\mathrm{h})=p_a(\beta_\mathrm{c})$. In this case, namely when $\epsilon_a=\tilde{\epsilon}$, the inequality \eqref{eq:crossing-inequality} becomes an equality and hence \eqref{eq:thermo-constraint-0} saturates. The energetic constraint can only be saturated if the internal energy of the system $U$ does not vary much in the considered interval, namely if $U(\beta_\mathrm{c})\approx U(\beta_\mathrm{h})$. If in addition, the energy of the initial state $\epsilon_a$ approximately equals this internal energy, the first term on the right hand side of Eq.~\eqref{eq:start-energetic} can be neglected and the energetic constraint \eqref{eq:en-constraint} saturates.

In contrast, the constraints are always trivially fulfilled when the function characterizing the response to a temperature variation, $\W^\text{(Resp)}_{a\to b}$ is negative, because the theremodynamic and energetic costs are positive by construction. This is the case, when the energy of the initial state $\epsilon_a$ differs from the internal energy of the system by more than the internal energy's standard deviation, see Appendix~\ref{app:negative-rate-response}. This means that the constraints are nontrivial only for those transition rates, where the initial state is ``typical" for the system, namely with an energy close to the internal energy of the system.

\subsection{Thermodynamic limit}\label{subsec:constraints-thermo-limit}

Here, we comment on the relevance of the constraints~\eqref{eq:thermo-constraint} and~\eqref{eq:en-constraint} when the system approaches the thermodynamic limit.
Notably, both the thermodynamic and the energetic constraints combine transition probabilities which are always smaller than one, with \emph{extensive} properties of the system, namely with its internal energy and with its nonequilibrium free energy. 
However, the extensive quantities only appear explicitly on the left-hand side of Eqs.~(\ref{eq:thermo-constraint}, \ref{eq:en-constraint}).
This feature implies that the left- and the right-hand sides behave very differently depending on the size of the system.
More concretely, if the extensive quantities in the thermodynamic limit scale as
\begin{equation}
\label{eq:extensive_scalings}
    U \to \lambda U,\qquad \Delta F^\text{(c,h)}\to \lambda \Delta F^\text{(c,h)}\ ,
\end{equation}
where $\lambda$ is the scale parameter, and the transition probabilities scale with an arbitrary scaling function $f(\lambda)$, i.e. $p_{a\to b} \to f(\lambda)p_{a\to b}$, both the thermodynamic constraint of Eq.~\eqref{eq:thermo-constraint} and the energetic constraint of Eq.~\eqref{eq:en-constraint} become trivial. 
Indeed, the right-hand side contribution from the response $\W^{(\text{Resp})}_{a\to b}$ is smaller than the standard deviation of the internal energy, as shown in Appendix~\ref{app:negative-rate-response}. Therefore, for $\lambda\to\infty$, the right-hand side becomes negligible and Eqs.~(\ref{eq:thermo-constraint}, \ref{eq:en-constraint}) then reduce to expressions that state the positivity of the thermodynamic and energetic costs, respectively, on the left-hand side.

However, the fact that the right-hand sides of these equations can be neglected in the thermodynamic limit---independently on how it is taken---leads to a trivial statement. The thermodynamic and energetic costs are positive by construction: The transition probabilities are positive $p_{a\to b}\geq 0$, and so are the coefficients of performance $\eta^{\text{(c,h)}}\geq0$, the nonequilibrium free energies $-\Delta F^{\text{(c,h)}}\geq 0$, as well as the energy variation $dU(x)$ in the integral in Eq.~\eqref{eq:en-constraint}.
Thus, both the thermodynamic and the energetic constraints pose constraints on small-scale quantum systems that do \textit{not} satisfy the thermodynamic limit.

\section{Particle and energy exchange in bipartite systems}\label{sec:bipartite}

Up to here, we have established and analyzed general constraints on transition probabilities in a two-point measurement scheme, without specifying how the transitions are induced.
One topic of interest in which occurring transitions are detected is, e.g., in bipartite systems exchanging energy and particles. This could be a system coupled to an environment or more generally arbitrary coupled systems. In the following, we will apply the thermodynamic and energetic constraints developed in Sec.~\ref{sec:transition-probabilities} on such a setting.
For simplicity, we focus on a regime where the coupling between the two subsystems is weak. This has the advantage that a temperature bias between the two subsystems can be clearly defined. Furthermore, direct relations between transitions rates and fluctuating transport quantities can be established, thereby exploiting our results to pose constraints between currents and noise in the presence of a large temperature bias.

\subsection{Weak tunnel coupling}\label{sec:theory}

We study a bipartite system with Hamiltonian $\hat{H}_0 = \hat{H}\L + \hat{H}\R$, where subsystems L (left) and R (right) may be taken as generic systems, possibly with strong interactions. Here, we take subsystem L as the system on which measurements are performed and subsystem R as the one inducing transition in subsystem L; this allows us to directly apply our results from Sec.~\ref{sec:transition-probabilities}.
The subsystems are coupled to each other by the tunneling Hamiltonian
\begin{equation}
\label{eq:V_Hamiltonian}
    \hat{V}(t) = \hat{A} e^{-i\omega t} + \hat{A}^\dagger e^{i\omega t},
\end{equation}
which we assume to be a small perturbation, and which induces transitions in the subsystems.
Further, in the weak tunnel-coupling regime, we describe the total system with the product state $\hat{\rho} = \hat{\rho}^{\mathrm{L}}\otimes\hat{\rho}^{\mathrm{R}}$, where $\hat{\rho}^{\alpha}$ is the Gibbs state, defined in Eq.~\eqref{eq:gibbs}, at inverse temperature $\beta_\alpha$.
With this, the rates $\Gamma_\rightleftarrows$ for absorbing or emitting a quantum of energy $\hbar\omega$,  induced by the coupling, are---in the spectral representation~\cite{Callen1951Jul,safi2011Nov}---given as \begin{equation}
\label{eq:rate-definition}
\begin{split}
    \Gamma_\rightarrow &\equiv \sum_{nm}\left[\frac{2\pi}{\hbar}\sum_{lk} |\A_{nmlk}|^2 \delta(\epsilon^\text{(L)}_{mn} + \epsilon^\text{(R)}_{kl} - \hbar\omega) p^\text{(R)}_l\right]p^\text{(L)}_n,\\
    \Gamma_\leftarrow  &\equiv \sum_{nm}\left[\frac{2\pi}{\hbar}\sum_{lk} |\A_{nmlk}|^2 \delta(\epsilon^\text{(L)}_{mn} + \epsilon^\text{(R)}_{kl} - \hbar\omega)p^\text{(R)}_k\right]p^\text{(L)}_m .
\end{split}
\end{equation}
See Appendix~\ref{app:weak-transition rates} for the derivation starting from two-point measurement transition probabilities. Here, $\A_{nmlk} \equiv \braket{mk|\hat{A}|nl}$ is the matrix element of $\hat{A}$ in the basis of the eigenstates of the generic Hamiltonian $\hat{H}_0$, namely $\hat{H}\L\ket{nl} = \epsilon^\text{(L)}_n\ket{nl},$ and $\hat{H}\R\ket{nl} = \epsilon^\text{(R)}_l\ket{nl}$.
Furthermore, we have defined the energy differences $\epsilon^{(\alpha)}_{mn} \equiv \epsilon^{(\alpha)}_{m} - \epsilon^{(\alpha)}_{n}$, and the occupation probability of the state $\ket{n}$ as $p^{(\alpha)}_n\ket{n} = \hat{\rho}^\alpha \ket{n}$.
The terms in the square brackets in Eq.~\eqref{eq:rate-definition}, which are multiplied by the initial probabilities $p_{n/m}^{(\mathrm{L})}$, represent the conditional probabilities per unit time for system L to transition from a given state $n$ to state $m$ ($\Gamma_\rightarrow$) with absorption of $\hbar\omega$ or from $m$ to $n$ ($\Gamma_\leftarrow$) with emission of $\hbar\omega$. We can thus, as presented in Sec.~\ref{sec:tunnel-constraints} below,  use the developed constraints for the transition rates $\Gamma_{\leftrightarrows}$, which are sums over two-point measurement transition \textit{rates}.
Note that the partition of the system into two weakly coupled subsystems is obviously not required for the general constraints of Eqs.~\eqref{eq:thermo-constraint} and \eqref{eq:en-constraint}, but importantly, it here allows us to implement a well-defined and meaningfull temperature bias. Indeed, the tunneling rates depend on two temperatures, $\Gamma_\leftrightarrows\equiv\Gamma_\leftrightarrows(\beta\L,\beta\R)$, via the occupation probabilities $p_n^{(\mathrm{L})}$ and $p_l^{(\mathrm{R})}$, see Eq.~\eqref{eq:rate-definition}.

Interestingly, the transition rates in the weak-tunneling regime can be directly connected to transport quantities, namely to a current and its zero-frequency noise~\cite{Callen1951Jul,Safi2014Jan}
\begin{subequations}\label{eq:current-noise-rates}
\begin{align}
    \frac{I}{q} &= \Gamma_\rightarrow - \Gamma_\leftarrow, \label{eq:current-rates}\\
    \frac{\S}{q^2} &= \Gamma_\rightarrow + \Gamma_\leftarrow,   \label{eq:noise-rates}
\end{align}
\end{subequations}
where $q$ is a generalized charge, defined in terms of an operator $\hat{Q}$ satisfying $[\hat{Q},\hat{H}_0]=0$ and $[\hat{Q},\hat{A}]=q\hat{A}$, 
see the derivation in Appendix~\ref{app:current_and_noise}.

\subsection{Constraints in the tunneling regime}\label{sec:tunnel-constraints}

The thermodynamic and energetic constraints on the nonequilibrium tunneling rates---in the presence of a temperature \textit{bias}---can be written starting from Eqs.~\eqref{eq:thermo-constraint} and \eqref{eq:en-constraint}, 
\begin{subequations}
\begin{eqnarray}\label{eq:thermodynamic-constraint}
    \W^\text{(Thermo)}_\rightleftarrows & \geq & \W^\text{(Resp)}_\rightleftarrows,\\
    \label{eq:energetic-constraint}
    \W^\text{(Energy)}_\rightleftarrows & \geq & \W^\text{(Resp)}_\rightleftarrows,
\end{eqnarray}
\end{subequations}
with the cost functions~\footnote{Notably, the derivatives in the rate response $\W^\text{(Resp)}_\rightleftarrows$ [Eq.~\eqref{eq:response-def}] can alternatively be understood in terms of higher-order correlation functions of the tunneling operator $\hat{A}$ in~\eqref{eq:V_Hamiltonian}. More specifically, we find that
$\partial\L\Gamma_\rightarrow = \frac{1}{\hbar^2}\int dt \langle \hat{A}_\text{H}^\dagger(t)e^{i\omega t}\hat{A} \,[U\L-\hat{H}\L]\rangle$, and equivalently for $\Gamma_\leftarrow$,
where we recall that $\hat{A}_\text{H}(0)= \hat{A}$.}
\begin{subequations}
\begin{eqnarray}
    \label{eq:response-def}
    \W^\mathrm{(Resp)}_\rightleftarrows  & \equiv & \partial\L \Gamma_\rightleftarrows (\beta_\mathrm{h},\beta_\mathrm{c}) - \partial\L \Gamma_\rightleftarrows(\beta_\mathrm{c},\beta_\mathrm{c}),\\\label{eq:thermo-def}
     \W^\text{(Thermo)}_\rightleftarrows &\equiv&  -\Delta F\L^{\text{(c)}} \eta^{\text{(h)}} \Gamma_\rightleftarrows(\beta_\mathrm{h},\beta_\mathrm{c}) \nonumber\\
    &\qquad& - \Delta F\L^{\text{(h)}}\eta^{\text{(c)}} \Gamma_\rightleftarrows(\beta_\mathrm{c},\beta_\mathrm{c}),\\\label{eq:en-def}
    \W^\text{(Energy)}_\rightleftarrows & \equiv & \int_{U\L(\beta_\mathrm{c})}^{U\L(\beta_\mathrm{h})}\!\!\!\! \Gamma_\rightleftarrows(x,\beta_\mathrm{c})d[U\L(x)]
\end{eqnarray}
\end{subequations}
Importantly, the rates here depend on two temperatures. We write this temperature dependence out explicitly, where the first argument is always the temperature of the left system and the second one the temperature of the right system. 
Note that we could have chosen \textit{any} temperature $T_\text{R}=\beta\R^{-1}$, as becomes clear from the derivation of the general constraints on transition probabilities in Sec.~\ref{sec:transition-probabilities}. Here, however, we choose one of the settings in the \textit{absence} of a temperature bias, where the two subsystems have equal temperatures, in order to be able to compare to an easily accessible, experimentally relevant reference situation when establishing  constraints for the out-of-equilibrium tunneling rate in the \textit{presence} of a temperature bias. We therefore deliberately choose $\beta\R\equiv\beta_\mathrm{c}$ and hence compare the tunneling rates when the two subsystems have the same temperature, i.e., $\Gamma_\rightleftarrows(\beta_\mathrm{c},\beta_\mathrm{c})$, with those under the desired out-of-equilibrium condition, i.e., $\Gamma_\rightleftarrows(\beta_\mathrm{h},\beta_\mathrm{c})$.
The rate response $\W^\mathrm{(Resp)}_\rightleftarrows$ accounts for both the equilibrium and out-of-equilibrium response of the tunneling rates to a change in the temperature of  subsystem L. Derivatives with respect to the first temperature argument are indicated in the cost function for the temperature response by $\partial_\mathrm{L}$. In what follows, for conciseness, we refer to the equal-temperature tunneling rates $\Gamma_\rightleftarrows(\beta_\mathrm{c},\beta_\mathrm{c})$ as \textit{equilibrium} tunneling rates, even though we want to emphasize here that the dependence on the energy transfer $\hbar\omega$, see Eq.~\eqref{eq:rate-definition}, implies the full treatment of possible nonequilibrium conditions beyond a temperature bias, induced by an external agent. 

We emphasize that Eq.~\eqref{eq:thermodynamic-constraint} directly translates into a constraint on the out-of-equilibrium noise, which is given by the sum of the tunneling rates via Eq.~\eqref{eq:noise-rates}. This also means, from a more practical, experimental viewpoint, that one possibility to test the constraint~\eqref{eq:thermodynamic-constraint} is by measuring the noise and tunneling current in two configurations: (i) When both subsystems have the same cold temperature, and (ii) in the desired out-of-equilibrium condition. One can then access the tunneling rates as $2q^2\Gamma_\rightleftarrows=\mathcal{S}\pm q I$.
All the other quantities entering the inequality can be calculated once the state of the subsystem considered at the different temperatures is known~\footnote{Note that it is \textit{not} necessary to know the Hamiltonian of the subsystem R kept at fixed temperature, which may be arbitrarily complicated, to test these constraints.}.
Similarly to the thermodynamic constraint, Eq.~\eqref{eq:energetic-constraint} directly translates into an energetic constraint on the integral over the out-of-equilibrium noise in Eq.~\eqref{eq:noise-rates} by taking the sum of the energetic constraints for both tunneling directions.
However, this integral makes the energetic constraint less relevant from an experimental point of view because it requires knowledge of both internal energy and tunneling rates at all intermediate temperatures of subsystem L.

From the thermodynamic constraint~\eqref{eq:thermodynamic-constraint}, we can also derive a direct lower bound on the out-of-equilibrium tunneling rates $\Gamma_\rightleftarrows(\beta_\mathrm{h},\beta_\mathrm{c})$ in terms of the \textit{equilibrium} rates (see the derivation in Appendix~\ref{app:Gronwall-inequality}).
These constraints furthermore contain the rate response, (integrals) of thermodynamic quantities, and they read
\begin{equation}\label{eq:gronwall_bound}
\begin{split}
    \Gamma_\rightleftarrows(\beta_\mathrm{h},\beta_\mathrm{c})&\geq \Gamma_\rightleftarrows(\beta_\mathrm{c},\beta_\mathrm{c})\exp\left[\int_{\beta_\mathrm{h}}^{\beta_\mathrm{c}} g(x)dx\right] + \\
    &\quad- \int_{\beta_\mathrm{h}}^{\beta_\mathrm{c}} f(x)\exp\left[\int_{\beta_\mathrm{h}}^x g(s)ds\right] dx,
\end{split}
\end{equation}
where we defined
\begin{equation}
    \begin{split}
 f(x) &\equiv \partial\L \Gamma_\rightleftarrows(\beta_\mathrm{c},\beta_\mathrm{c}) - \Delta F\L^\text{(c)}(x)\eta^\text{(h)}(x) \Gamma_\rightleftarrows(\beta_\mathrm{c},\beta_\mathrm{c}),\\ 
 g(x) &\equiv -\Delta F\L^\text{(h)}(x)\eta^\text{(c)}(x).
    \end{split}
\end{equation}
While the bound~\eqref{eq:gronwall_bound} has a more complex shape, containing integrals over thermodynamic functions, it has the important advantage that it does not depend on the out-of-equilibrium responses $\partial\L\Gamma_\rightleftarrows(\beta_\mathrm{h},\beta_\mathrm{c})$. Instead, only the more easily accessible \textit{equilibrium response function}, $\partial\L \Gamma_\rightleftarrows(\beta_\mathrm{c},\beta_\mathrm{c})$, enters Eq.~\eqref{eq:gronwall_bound}.
We emphasize that Eq.~\eqref{eq:gronwall_bound} directly provides a lower bound for the out-of-equilibrium noise.
In particular, compared to~\cite{Safi2014Jan}, it provides a nontrivial constraint indicating \textit{how much} super-Poissonian the noise is.

A further insightful way of writing  the constraints of Eqs.~(\ref{eq:thermodynamic-constraint}, \ref{eq:energetic-constraint}), is by highlighting their contributions from sums over resonances. Indeed, we notice that both the thermodynamic, $\W_\rightleftarrows^\text{(Thermo)}$, and energetic cost, $ \W_\rightleftarrows^\text{(Energy)}$, as well as the rate response $\W_\rightleftarrows^\text{(Resp)}$
can be recast as 
\begin{equation}
    \W^{(i)}_\leftrightarrows = \sum_{nmlk} w^{(i)}_{\leftrightarrows, nmlk} \delta(\epsilon_{mn}^\text{(L)}+ \epsilon_{kl}^\text{(R)} -\hbar\omega),
\end{equation}
where $i\in\{\text{Thermo, Energy, Resp}\}$.
This is done by using the expression of the rates in Eq.~\eqref{eq:rate-definition} and the definitions in Eqs.~(\ref{eq:response-def}, \ref{eq:thermo-def}, \ref{eq:en-def}). For the examples studied below, we focus on the amplitude of the resonance at energy $\hbar\omega$ of interest by considering
\begin{equation}
\label{eq:Ci_coeff}
    \mathcal{C}_\rightleftarrows^{(i)}(\omega) \equiv \sum_{\{nmlk|\epsilon_{mn}^\text{(L)}+ \epsilon_{kl}^\text{(R)} =\hbar\omega\}} w^{(i)}_{\rightleftarrows, nmlk},
\end{equation}
and thus sum over all the resonances at the same frequency $\omega$.
Since both the thermodynamic and energetic constraints hold separately at each resonance, Eqs.~(\ref{eq:thermodynamic-constraint}, \ref{eq:energetic-constraint})  also hold for each amplitude $\mathcal{C}_\rightleftarrows^{(i)}(\omega)$, i.e.
\begin{equation}
\label{eq:constraints_C}
    \begin{split}
        \mathcal{C}_\rightleftarrows^\text{(Thermo)}(\omega)&\geq \mathcal{C}_\rightleftarrows^\text{(Resp)}(\omega),\\
        \mathcal{C}_\rightleftarrows^\text{(Energy)}(\omega)&\geq \mathcal{C}_\rightleftarrows^\text{(Resp)}(\omega).\\
    \end{split}
\end{equation}
This establishes natural quantities, which we will compute in the following  analysis of the example systems.
\begin{figure}
    \centering
    \includegraphics{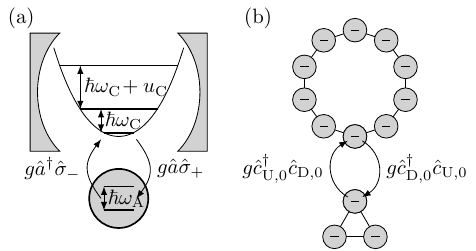}
    \caption{Illustration of the two setups considered in Sec.~\ref{sec:bipartite}. (a) A two-level atom with frequency $\omega_\text{A}$ is weakly coupled, $g\ll \hbar\omega_\text{A,C}$, to a cavity with characteristic frequency $\omega_\text{C}$ and Kerr non-linearity $u_\mathrm{C}$, see Sec.~\ref{subsec:Jaynes-Cummings}. (b) Two interacting fermionic tight-binding rings with different sizes are weakly coupled at a single (``0'') site, see Sec.~\ref{subsec:fermionic-rings}.}
    \label{fig:examples-sketch}
\end{figure}

Concretely, to illustrate the constraints \eqref{eq:thermodynamic-constraint} and \eqref{eq:energetic-constraint}, we consider two different physical settings, represented by the systems depicted in Fig.~\ref{fig:examples-sketch}: An atom coupled to a nonlinear cavity [panel (a)], see Sec.~\ref{subsec:Jaynes-Cummings}, and two fermionic chains interchanging particles at a single site [panel (b)], see Sec.~\ref{subsec:fermionic-rings}.
Such examples are chosen not only because of illustration purposes, but also because they are of experimental relevance to test our predictions with state-of-the art setups.
These two examples include optical or mechanical cavities coupled together~\cite{Leijssen2017Jul} or to (artificial) atoms~\cite{Leibfried2003Mar,Kaufman2012Nov,Valmorra2021Sep,Vigneau2022Dec,Haldar2024Feb, Sundelin2024Mar}, and tunneling bridges across molecules~\cite{Pyurbeeva2021Nov,Gehring2021Apr,Gemma2023Jun} or magnetic impurities~\cite{Hsu2022Apr, Thupakula2022Jun, Trishin2023Oct}.

\subsection{Atom coupled to nonlinear cavity}\label{subsec:Jaynes-Cummings}

\begin{figure}
    \centering
    \includegraphics{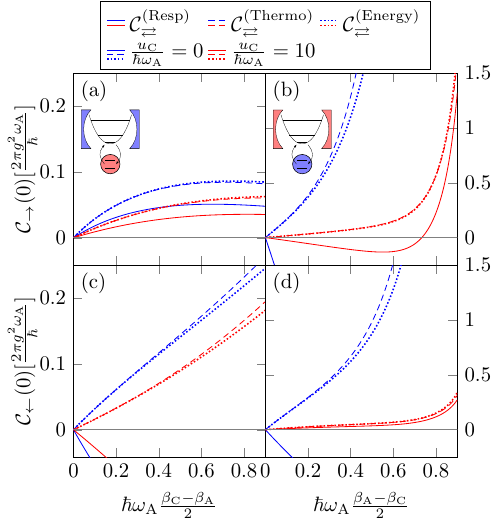}
    \caption{Thermodynamic cost (dashed lines), energetic cost (dotted lines), and rate response (solid lines) for an atom coupled to a cavity as functions of the inverse temperature difference $\beta_\text{A} - \beta_\text{C}$. The atom and cavity frequencies are taken in resonance, $\omega_\text{A}=\omega_\text{C}$, and the different curve colors denote different Kerr nonlinearities, $u_\text{C}/\hbar\omega_\text{A}=0,10$, in blue and red, respectively. In panels (a, c) the cavity is kept at the cold temperature ($\beta_\text{C}>\beta_\text{A}$), whereas in panels (b, d) it is the atom that has the colder temperature ($\beta_\text{A}>\beta_\text{C}$).}
    \label{fig:Jaynes-Cummings}
\end{figure}

In this section, we consider an atom weakly coupled to a nonlinear cavity~\cite{Werner1991Oct}. Its Hamiltonian is given by
\begin{equation}\label{eq:Jaynes-Cummings-Hamiltonian}
    \hat{H} = \hbar\omega_\text{A}\frac{\hat{\sigma}_z}{2} + \hbar\omega_\text{C} \hat{n} + \frac{u_\text{C}}{2}\hat{n}(\hat{n}-1) + g(\hat{a} \hat{\sigma}_+ + \hat{a}^\dagger \hat{\sigma}_-).
\end{equation}
Here, one of the subsystems is the atom, described by the Pauli matrix $\hat{\sigma}_z$ and characterized by the frequency $\omega_\text{A}$.
The cavity is the other subsystem, and is described by the number operator $\hat{n} = \hat{a}^\dagger \hat{a}$, with $[\hat{a},\hat{a}^\dagger]=1$, and the cavity frequency $\omega_\text{C}$.
We also include a Kerr nonlinearity, parametrized by $u_\text{C}$, which plays the role of effective interactions between cavity photons in this system.
The atom and the cavity exchange photon quanta through the weak-tunneling term $g(\hat{a} \hat{\sigma}_+ + \hat{a}^\dagger \hat{\sigma}_-)$ with $g\ll \hbar\omega_\text{A,C}$ and where $\hat{\sigma}_+$ and $\hat{\sigma}_-$ are the raising and lowering operators of the atom states.
Note that, in the language of Eq.~\eqref{eq:V_Hamiltonian}, we choose $\omega=0$ since any external driving frequency can be incorporated in the laser detuning---replacing $\omega_\mathrm{C}$---in the rotating frame~\cite{Aspelmeyer2014Dec}.
For $u_\text{C}=0$, the Hamiltonian~\eqref{eq:Jaynes-Cummings-Hamiltonian} reduces to the Jaynes-Cummings Hamiltonian~\cite{Jaynes1963Jan}.

To relate to the theoretical framework of Sec.~\ref{sec:theory}, we now identify the tunneling operator $\hat{A}=g\hat{a} \hat{\sigma}_+$, which we use to calculate the zero-frequency amplitudes $\mathcal{C}_\rightleftarrows^{(i)}(0)$ in Eq.~\eqref{eq:Ci_coeff}.
Note that a possible observable $\hat{Q}$ with which a current and its noise may be defined according to~\eqref{eq:current-noise-rates} could here be the atomic occupation $\hat{Q} = \frac{\hat{\sigma}_z+1}{2}$, satisfying $[\hat{Q}, \hat{A}]=\hat{A}$.
The result for $\mathcal{C}_\rightleftarrows^{(i)}(0)$ is shown in Fig.~\ref{fig:Jaynes-Cummings}, where we plot $\mathcal{C}_\rightleftarrows^{(i)}(0)$ for different values of the non-linearity $u_\text{C}$.
Since we are free to choose the L subsystem, in panels (a,~c) the atom is the subsystem considered at two different temperatures, whereas in panels (b,~d) the cavity is.
In general, Fig.~\ref{fig:Jaynes-Cummings} shows that the thermodynamic cost and the energetic cost are very similar to each other and that there is no hierarchy between them, namely it depends on the specific parameters whether the thermodynamic or the energetic cost is larger~\footnote{ This can for example be seen in panel (b), where, in the absence of Kerr non-linearity the thermodynamic cost is larger than the energetic one, but in the presence of such a non-linearity the opposite holds true.}. We furthermore see that---unsurprisingly---both thermodynamic and energetic constraints are trivially saturated at equal temperature.
However, for sizable temperature biases, considering the atom or the cavity at different temperatures affects the constraints.

Panels (a,~c) illustrate the case where the cavity is always taken at the cold temperature, while we compare the rates for the atom being taken at two different temperatures. 
We see that the rate $\Gamma_\rightarrow$ has a nontrivial constraint via the thermodynamic and energetic cost: for all temperature biases,  the response contribution $\mathcal{C}^{(\mathrm{Resp})}_\rightarrow$ is positive and it thereby constrains the thermodynamic and energetic costs non-trivially. In contrast, for the rate $\Gamma_\leftarrow$, the response contribution is always negative, thereby not putting any nontrivial constraint on the energetic and the thermodynamic cost, which are positive by definition.
This can be understood from the fact that the rate $\Gamma_\leftarrow$ characterizes the process in which the atom emits a photon into the cavity, requiring the atom to be in an excited state. However, if all allowed transitions start in states with energies that are more than a standard deviation away from the internal energy of the subsystem, here the atom, then the rate response is negative $\W^\text{(Resp)}\leq 0$, and the constraints become trivial, see Appendix~\ref{app:negative-rate-response} for a detailed proof.

In panels (b,~d), we show results where the atom is always kept at the cold temperature, while the cavity can be taken at two different temperatures. Here, we see that the bounds are trivial in the absence of the non-linearity, $u_\mathrm{C}=0$. Instead, a nonvanishing cavity non-linearity allows one to approach both thermodynamic and energetic constraints also at large temperature biases for both rates $\Gamma_\rightleftarrows$.
This can be understood from how the nonlinearity affects the constraints~\eqref{eq:constraints_C}. There are two effects of the nonlinearity: On the one hand, finite values of $u_\text{C}$ break the degeneracy of the atom-cavity transitions, making the atom couple only to two consecutive cavity states. On the other hand, a large $u_\text{C}$ increases the energy spacing of the cavity. This feature reduces the number of states that have non negligible occupation at finite cavity temperature $\beta_\text{C}\neq 0$. 
These two aspects effectively reduce the  size of the cavity (which here plays the role of the system on which the two-point measurement is performed, see Sec.~\ref{sec:transition-probabilities}), moving it further away from the thermodynamic limit and thereby making the bounds more constraining.

\subsection{Coupled fermionic rings}\label{subsec:fermionic-rings}

\begin{figure}
    \centering
    \includegraphics{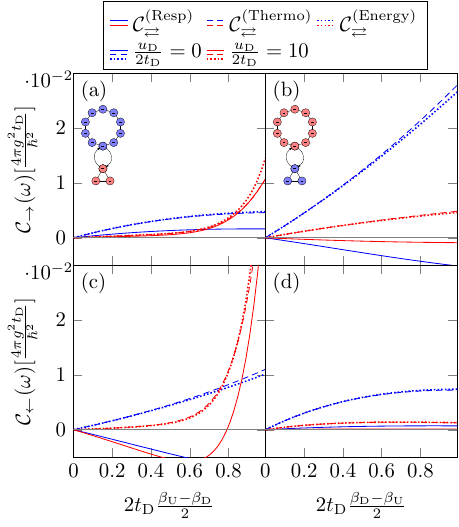}
    \caption{Thermodynamic cost (dashed lines), energetic cost (dotted lines), and rate response (solid lines) for two (U and D) coupled fermionic rings with sizes $L_\text{D}=3$ and  $L_\text{U}=10$. The costs are plotted vs the inverse temperature difference $\beta_\text{D} - \beta_\text{U}$.
    The ring hopping parameters, $t_\text{D}$ and $t_\text{U}$, and the driving frequency are fixed as $t_\text{U}=\frac{4}{5}t_\text{D} = 2\hbar\omega$. The upper ring charging energy $u_\text{U}=0$ is also fixed, while the down ring charging energy is taken as $u_\text{D}/2t_\text{D} = 0,10$ for curves in blue and red, respectively. In panels (a,~c) the upper, larger ring is kept at the cold temperature ($\beta_\text{U}>\beta_\text{D}$), whereas in panels (b,~d) it instead is the down, smaller ring that is colder ($\beta_\text{D}>\beta_\text{U}$).}
    \label{fig:fermionic-rings-interaction}
\end{figure}

\begin{figure}
    \centering
    \includegraphics{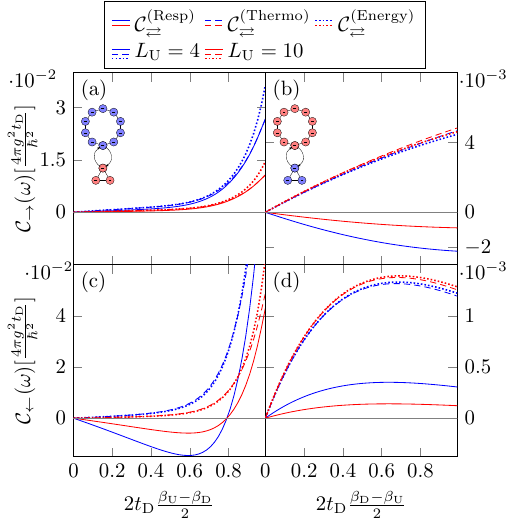}
    \caption{Thermodynamic cost (dashed lines), energetic cost (dotted lines), and rate response (solid lines) for the two coupled fermionic rings with charging energies $u_\text{D}/2t_\text{D}=10, u_\text{U}=0$, plotted vs the inverse temperature difference $\beta_\text{U} - \beta_\text{D}$.
    The lower ring has fixed size $L_\mathrm{D}=3$, and the upper ring has sizes $L_\text{U}=4,10$ for the curves in blue and red, respectively.
    All other parameters are the same as in Fig.~\ref{fig:fermionic-rings-interaction}.
    In panels (a,~c), the upper, larger ring is kept at the cold temperature ($\beta_\text{U}>\beta_\text{D}$), whereas in panels (b,~d) the lower, smaller ring is kept fixed at the cold temperature ($\beta_\text{D}>\beta_\text{U}$).}
    \label{fig:fermionic-rings-size}
\end{figure}

Here, we consider two  fermionic tight-binding rings with $L_\alpha$ sites for subsystem $\alpha=\text{D, U}$, see Fig.~\ref{fig:examples-sketch}. Their Hamiltonians read
\begin{equation}
\label{eq:fermionic_rings}
\begin{split}    
    \hat{H}_\alpha &= \sum_{i=0}^{L_\alpha-1}t_\alpha(\hat{c}_{\alpha, i+1}^\dagger \hat{c}_{\alpha, i}+\hat{c}_{\alpha, i}^\dagger \hat{c}_{\alpha, i+1})+ \frac{u_\alpha}{2} \hat{N}_\alpha(\hat{N}_\alpha-1),
\end{split}
\end{equation}
where we take periodic boundary conditions $\hat{c}_{\alpha, L_\alpha} = \hat{c}_{\alpha, 0}$ for the fermionic operators obeying $\{\hat{c}_{\alpha,i}, \hat{c}_{\beta,j}^\dagger\} = \delta_{\alpha\beta}\delta_{ij}$. The charging energy contribution in Eq.~\eqref{eq:fermionic_rings}, parametrized by $u_\alpha$, depends on $\hat{N}_{\alpha} = \sum_{i=0}^{L_\alpha-1} \hat{c}_{\alpha, i}^\dagger \hat{c}_{\alpha, i}$, i.e., the total number operator for subsystem $\alpha$.
Next, we introduce weak tunneling between the two rings, taken at the site $i=0$, by adding the tunneling Hamiltonian
\begin{equation}
    \hat{V}(t) = g\left( \hat{c}^\dagger_{\text{D},0}\hat{c}_{\text{U},0}e^{-i\omega t} + \hat{c}^\dagger_{\text{U},0}\hat{c}_{\text{D},0}e^{i\omega t}\right),
\end{equation}
with $g\ll t_\alpha$.
Here, we identify the tunneling operator $\hat{A} = g \hat{c}^\dagger_{\text{D},0} \hat{c}_{\text{U},0}$, which transfers a fermion from the upper ring (U) to the lower ring (D).
Then, using this tunneling operator $\hat{A}$, we calculate the amplitudes $\mathcal{C}_\rightleftarrows^{(i)}(\omega)$ of Eq.~\eqref{eq:Ci_coeff}. 
Note that a possible observable $\hat{Q}$ could here be the number of fermions in one ring, e.g. $\hat{Q}=\hat{N}_\text{D}$, which satisfies $[\hat{Q}, \hat{A}]=\hat{A}$.
The result for $\mathcal{C}_\rightleftarrows^{(i)}(\omega)$ is shown in Fig.~\ref{fig:fermionic-rings-interaction}, where we plot $\mathcal{C}_\rightleftarrows^{(i)}(\omega)$ at different values of the charging energy  $u_\text{D}$ in the lower ring, and in Fig.~\ref{fig:fermionic-rings-size} for different sizes $L_\text{U}$ of the upper ring.
Note that the frequency $\omega$ of the tunneling Hamiltonian can emerge from a potential bias between the rings after a gauge transformation~\cite{Rogovin1974Jul,Safi2010}.

Similarly to the atom-cavity system considered in Sec.~\ref{subsec:Jaynes-Cummings}, the charging energy $u_\text{D}$ influences the thermodynamic and energetic constraints in the system of coupled rings. In panels (a,~c) of Fig.~\ref{fig:fermionic-rings-interaction}, we show results where the lower ring, namely the one displaying finite interaction, is the one that is considered at two different temperatures when comparing out-of-equilibrium rates. This produces nontrivial constraints for both $\Gamma_\rightleftarrows$, similar to the case of the atom-cavity system, where the cavity with nonlinearity $u_\mathrm{C}$ is taken at two different temperatures, as shown in Fig.~\ref{fig:Jaynes-Cummings}~(b,~d). 
In contrast, as shown in panels (b,~d) of Fig.~\ref{fig:fermionic-rings-interaction}, for the case where the upper ring is taken at two different temperatures, only the rate $\Gamma_\leftarrow$ is non-trivially constrained, as long as the charging energy $u_\text{U}=0$ vanishes.

Furthermore, as discussed in Sec.~\ref{subsec:constraints-thermo-limit}, increasing the size of the ring that is considered at two different temperatures (namely corresponding to the one on which the two-point measurement is performed, see Sec.~\ref{sec:transition-probabilities}), when comparing out-of-equilibrium rates, weakens the constraints, as is seen in panels (b,~d) of Fig.~\ref{fig:fermionic-rings-size}.
In this example, the tunneling rates scale inversely with the ring size, i.e. with a scaling function $f$ on the form $f(L_\text{U})\sim1/L_\text{U}$ [see below Eq.~\eqref{eq:extensive_scalings}], whereas the extensive quantities entering the thermodynamic and energetic costs scale linearly with $L_\text{U}$.
Thus, while the costs are essentially unaffected upon increasing $L_\text{U}$, the rate responses instead decrease.
By contrast, if the ring that is considered at two different temperatures has a fixed size, both costs and rate responses scale as $1/L_\text{U}$ because of the scaling of the tunneling rates alone, see panels (a,~c). This feature implies that, as long as the size of the subsystem considered at two different temperatures remains small, there are nontrivial constraints on the tunneling rates, irrespective of the size of the subsystem with fixed temperature.

\section{Conclusions}\label{sec:conclusions}

We have studied the transition probabilities of two-point measurement schemes for different initial thermal state occupations, in an otherwise generic quantum system.
For such setups, we have proved two novel bounds that the transition probabilities~\eqref{eq:transition_probability} satisfy: i) The thermodynamic constraint~\eqref{eq:thermo-constraint}, stating that the response of the transition probabilities for different initial temperatures bounds from below the magnitude of the probabilities themselves multiplied by the heat dissipated while cooling or heating the system to these temperatures. ii) The energetic constraint~\eqref{eq:en-constraint}, which states that the response of the transition probabilities to temperature variations also bounds from below the transition probabilities weighted by the continuous change in the internal energy required to heat the system.

As one application of interest, we have analyzed the developed constraints for the tunneling rates of transitions between two weakly coupled subsystems driven out of equilibrium by a temperature bias. 
As a key consequence of these bounds, also the tunneling current and its low-frequency noise become bounded in the presence of a temperature bias. 

Our results thus highlight a fundamental connection between thermodynamic potentials and transport quantities for small-size quantum systems. In particular, they should be testable for a broad range of state-of-the-art experimental setups, including optical or mechanical cavities coupled together~\cite{Leijssen2017Jul} or to (artificial) atoms~\cite{Leibfried2003Mar,Kaufman2012Nov,Valmorra2021Sep,Vigneau2022Dec,Haldar2024Feb, Sundelin2024Mar}, and tunneling bridges across molecules~\cite{Pyurbeeva2021Nov,Gehring2021Apr,Gemma2023Jun} or magnetic impurities~\cite{Hsu2022Apr, Thupakula2022Jun, Trishin2023Oct}.
Beyond the weak-coupling examples studied in the present paper, also the strong coupling regime is captured by our general constraints and it would be intriguing to investigate, e.g., coherent oscillations between strongly coupled few-level systems in the future.

\acknowledgments
We gratefully acknowledge funding from the Knut and Alice Wallenberg foundation via the fellowship program (L.T. and J.S.), the European Research Council (ERC) under the European Union’s Horizon Europe research and innovation program (101088169/NanoRecycle) (J.S.), from the PNRR MUR project No. PE0000023-NQSTI (M.A.), from the Swedish Vetenskapsrådet via Project No. 2023-04043 (C.S.), and from the European Union's Horizon 2020 research and innovation programme under grant agreement No 101031655 (TEAPOT) (C.S.). 

\appendix

\section{Work extraction and nonequilibrium free energy}\label{app:extracting-work}
By closely following Ref.~\cite{Esposito2010Jan}, we show in this Appendix how the nonequilibrium free energy limits work extraction.
Consider a system S coupled to a bath B, described by the total Hamiltonian
\begin{equation}
    \hat{H}(t) = \hat{H}\Sys(t) + \hat{H}\B + \hat{H}_\text{SB}(t).
\end{equation}
Here, $\hat{H}\Sys(t)$ and $\hat{H}\B$ is the system and bath Hamiltonian, respectively, and $\hat{H}_\text{SB}(t)$ is the time-dependent coupling between system and bath. 
The rate of work done on the system and the heat production in the bath are defined as
\begin{subequations}
\begin{align}
    \dot{W}(t) &\equiv \Tr{}{\frac{d \hat{H}(t)}{dt}\hat{\rho}(t)},\\
    \dot{Q}\B(t) &\equiv \Tr{}{\hat{H}\B \frac{d\hat{\rho}(t)}{dt}} \\
    &= -\Tr{}{\left(\hat{H}\Sys(t) + \hat{H}_\text{SB}(t)\right)\frac{d\hat{\rho}(t)}{dt}}.\nonumber
\end{align}
\end{subequations}
In this way, we can write an expression equivalent to the first law of thermodynamics. To this end, one considers the change in the energy of S, denoted $U(t) \equiv \Tr{}{\left(\hat{H}\Sys(t) + \hat{H}_\text{SB}(t)\right)\hat{\rho}(t)}$, namely
\begin{equation}
    \frac{d U(t)}{dt} = \dot{W}(t) - \dot{Q}\B(t).
\end{equation}
Integrating from the beginning of the work extraction operation at $t=0$ to its end at $t=\tau$, we have
\begin{equation}\label{app:eq:1st-law}
    \Delta U(\tau)\equiv U(\tau) - U(0) =  W(\tau) - \Delta Q\B(\tau).
\end{equation}
To write an expression equivalent to the second law of thermodynamics, we assume that before the work extraction, the system and the bath are uncorrelated, i.e. $\hat{\rho}(0)=\hat{\rho}\Sys(0)\otimes \hat{\rho}\B(0)$.
Then, the change in entropy of system and bath can be written in terms of the relative entropy between the initial and the time-evolved density matrices. Generally, the relative entropy between two density matrices $\hat{\rho}_1$ and $\hat{\rho}_2$ is defined as
\begin{align}
   D[\hat{\rho}_1||\hat{\rho}_2]\equiv \Tr{}{\hat{\rho}_1 (\ln\hat{\rho}_1 - \ln{\hat{\rho}_2})}, 
\end{align}
which applied to the evolution of the coupled system and bath becomes
\begin{equation}\label{app:eq:2nd-law}
\begin{split}
    D[\hat{\rho}(t)||\hat{\rho}\Sys(t)\otimes \hat{\rho}\B(t)] &= \Delta S\Sys(t) + \Delta S\B(t) \geq 0.
\end{split}
\end{equation}
Here, $\Delta S_\alpha(t)\equiv S[\hat{\rho}_\alpha(t)] - S[\hat{\rho}_\alpha(0)]$ is the difference in von Neumann entropy, and the positivity stems from Klein's inequality.
If the bath is sufficiently large, its entropy change is well approximated by the Clausius relation
\begin{equation}\label{app:eq:Clausius-relation}
    \Delta S\B(t) \approx  \frac{\Delta Q\B(t)}{T\B},
\end{equation}
where $T\B$ is the temperature of the bath.
Then, combining Eqs.~(\ref{app:eq:1st-law}, \ref{app:eq:2nd-law}, \ref{app:eq:Clausius-relation}) we find that the performed work is limited from below by the nonequilibrium free energy $\Delta F$ as
\begin{equation}
    W(\tau) \geq \Delta U(\tau) - T\B\Delta S\Sys(\tau) \equiv \Delta F.
\end{equation}
In the present paper, we are interested in the situation where, at the beginning and at the end of the operation, the system Hamiltonian and the system-bath interaction satisfy
\begin{equation}
    \hat{H}\Sys(0)=\hat{H}\Sys(\tau),\quad \hat{H}_\text{SB}(0)=\hat{H}_\text{SB}(\tau)=0,
\end{equation}
i.e., the system Hamiltonian is left unchanged after the operation, and no system-bath coupling exists at the beginning and at the end of the operation. Then, the nonequilibrium free energy reads
\begin{equation}
    \Delta F = \Delta F\Sys = \Delta U\Sys(\tau) - T\B\Delta S\Sys(\tau),
\end{equation}
with $\Delta U\Sys(\tau) \equiv U\Sys(\tau) - U\Sys(0)$ and $U\Sys(t) \equiv \Tr{\mathrm{S}}{\hat{H}\Sys(t)\hat{\rho}\Sys(t)}$. These features allow us to calculate the maximum work that can be extracted $W^\text{ext}(\tau) = -W(\tau)$ by using only the knowledge of the system's density matrix at the beginning and at the end of the operation:
\begin{equation}
    W^\text{ext}(\tau) \leq - \Delta F\Sys.
\end{equation}

Furthermore, if the initial and final states of the system are thermal states at temperatures $T$ and $T_\text{B}$, respectively, the entropy variation of the system reads
\begin{equation}\label{app:eq:entropy-production-between-thermal-states}
    \Delta S\Sys(\tau) = \beta \Delta U\Sys(\tau) + (\beta\B-\beta)U\Sys(\tau) + \ln\left(\frac{Z\Sys(\tau)}{Z\Sys(0)}\right),
\end{equation}
with $Z\Sys(0)=\Tr{\mathrm{S}}{\exp(-\beta \hat{H}\Sys(0))}$ and $Z\Sys(\tau)=\Tr{\mathrm{S}}{\exp(-\beta\B \hat{H}\Sys(\tau))}$ being the partition functions at the inverse temperatures $\beta = T^{-1}$ and $\beta\B= T\B^{-1}$, respectively.
We now recall that the internal energy of a thermal state is related to the partition function as $U\Sys(\tau) = -\frac{\partial}{\partial \beta\B}\ln Z\Sys(\tau)$, and by using the concavity of $-\ln Z\Sys$ as a function of the inverse temperature, Eq.~\eqref{app:eq:entropy-production-between-thermal-states} leads to the inequality
\begin{equation}\label{app:eq:entropy-production-inequality}
    \Delta S\Sys(\tau) \leq \beta \Delta U\Sys(\tau).
\end{equation}
We next combine Eqs.~(\ref{app:eq:1st-law}, \ref{app:eq:2nd-law}, \ref{app:eq:entropy-production-inequality}) in two different ways, depending on whether we focus on the heat absorbed by the bath, $\Delta Q\B(\tau)$, or on the energy variation in the system $\Delta U\Sys(\tau)$. For these two situations, we find
\begin{subequations}
    \begin{align}
        W(\tau) - \left(1- \frac{T\B}{T}\right) \Delta U\Sys(\tau) \geq 0, \label{app:eq:work-and-internal-energy}\\
        W(\tau) - \left(1- \frac{T}{T\B}\right) \Delta Q\B(\tau) \geq 0. \label{app:eq:work-and-bath}
    \end{align}
\end{subequations}
As detailed in Fig.~\ref{fig:thermo-heat-cool}, we are interested in both cases $T<T\B$ and $T>T\B$, depending on whether we are cooling or heating the system.
When $T<T\B$, we focus on Eq.~\eqref{app:eq:work-and-internal-energy}. There, $\Delta U\Sys(\tau)\geq 0$, since the system is heated by the hot bath, and $W(\tau)\leq 0$ since we are using the heat flow to extract energy. Then
\begin{equation}
    \Delta U\Sys(\tau) \geq W^\text{ext}(\tau) \frac{T}{T\B-T} = W^\text{ext}(\tau) \eta^\text{(c)}.
\end{equation}
Instead, when $T>T\B$, we focus on Eq.~\eqref{app:eq:work-and-bath}, where $\Delta Q\B(\tau)\geq 0$ as the cold bath receives heat from the (initially hotter) system, and again $W(\tau)\leq 0$ as we are extracting work. Then,
\begin{equation}
    \Delta Q\B(\tau) \geq W^\text{ext}(\tau) \frac{T\B}{T\B-T} = W^\text{ext}(\tau) \eta^\text{(h)}.
\end{equation}

\section{Sufficient condition for trivial constraints}\label{app:negative-rate-response}

Both thermodynamic and energetic costs, Eqs.~\eqref{eq:thermo-term} and \eqref{eq:en-term} respectively, are positive by definition, and are lower-bounded by the response term $\W^\text{(Resp)}_{a\to b}$ in Eq.~\eqref{eq:response-term}.
Therefore, the thermodynamic and energetic constraints are nontrivial whenever $\W^\text{(Resp)}_{a\to b}> 0$.
In this Appendix we provide a sufficient condition for $\W^\text{(Resp)}_{a\to b}< 0$, making the results of Eqs.~(\ref{eq:thermo-constraint}, \ref{eq:en-constraint}) trivial, and a sufficient condition for $\W^\text{(Resp)}_{a\to b}> 0$, making the results of Eqs.~(\ref{eq:thermo-constraint}, \ref{eq:en-constraint}) nontrivial.
Starting from the definition of the response contribution~\eqref{eq:response-term} and recalling that derivatives with respect to $\beta$ only act on the probabilities $p_a$, we consider
\begin{equation}
    \partial_\beta p_a (\beta) = \left[U(\beta) - \epsilon_a\right]p_a(\beta).
\end{equation}
Taking a second derivative with respect to $\beta$ we find
\begin{equation}
    \partial_\beta^2 p_a(\beta) = \left[\partial_\beta U(\beta)+ [U(\beta)-\epsilon_a]^2 \right]p_a(\beta).
\end{equation}
We now focus on all energies $\epsilon_a$ lying far away from the internal energy $U$. Namely, we suppose that the distance between $\epsilon_a$ and $U$ is larger than the standard deviation of the internal energy, such that the following inequality holds for all $\beta\in[\beta_\mathrm{h}, \beta_{\mathrm{c}}]$,
\begin{equation}
        |U(\beta)-\epsilon_a| \geq  \delta U(\beta),\label{eq:interval}
\end{equation}
where 
\begin{equation}
   \delta U(\beta) = \sqrt{\sum_a (\epsilon_a)^2 p_a(\beta) - U^2(\beta)} = \sqrt{-\partial_\beta U(\beta)}
\end{equation} 
is the standard deviation of the internal energy.
In this energy interval, defined by Eq.~\eqref{eq:interval}, the second derivative of the occupation probabilities is positive, i.e. $\partial^2_\beta p_a(\beta)\geq 0$, making $\partial_\beta p_a(\beta)$ an increasing function in the inverse temperature $\beta$.
This in turn makes the rate response negative, 
\begin{equation}
    \partial_\beta p_a(\beta_\text{h}) -\partial_\beta p_a(\beta_\text{c}) \leq 0\quad\Rightarrow\quad \W^\text{(Resp)}_{a\to b}\leq 0,
\end{equation}
thus trivializing both thermodynamic and energetic constraints.

Conversely, the condition that the energies of the initial states lie \textit{within} a standard deviation from the internal energy for all inverse temperatures in $[\beta_\text{h}, \beta_\text{c}]$ is a sufficient condition for the rate response to be positive $\W^\text{(Resp)}_{a\to b}\geq 0$. 
Indeed, this condition on the energies makes $\partial_\beta p_a(\beta)$ a decreasing function in the inverse temperature $\beta$, leading to
\begin{equation}
      \partial_\beta p_a(\beta_\text{h}) -\partial_\beta p_a(\beta_\text{c})\geq\, 0\quad\Rightarrow\quad \W^\text{(Resp)}_{a\to b}\geq 0.
\end{equation}

We demonstrate this reasoning with a simple example: for the two-level system considered in Sec.~\ref{subsec:Jaynes-Cummings}, we can easily compute both the internal energy and its standard deviation
\begin{equation}
\begin{split}
    U_\text{A}(\beta) &= \hbar\omega_\text{A} p_1^\text{(A)}\\
    \delta U_\text{A}(\beta)&=\hbar\omega_\text{A} \sqrt{p_1^\text{(A)}[1-p_1^\text{(A)}]}
\end{split}
\end{equation}
with $p_1^\text{(A)}=\left(1+e^{\beta\hbar\omega_\text{A}}\right)^{-1}\leq 1/2$.
Therefore, when only the transition in which the atom \textit{emits} a photon into the cavity is allowed, the only possible initial state for the atom is the excited one, with energy $\hbar\omega_\text{A}$, which however satisfies
\begin{equation}
    \hbar\omega_\text{A} \geq  U_\text{A}(\beta) +  \delta U_\text{A}(\beta)
\end{equation}
for all $\beta$. Therefore, the rate response associated with the emission of a photon is always negative, and it does neither put a constraint on the thermodynamic cost nor on the energetic cost.

\section{Transition rates in the weak-tunneling regime}\label{app:weak-transition rates}

In this Appendix, we derive expressions for the weak-tunneling transition rates of Eq.~\eqref{eq:rate-definition} starting from the two-point measurement probabilities of Eq.~\eqref{eq:transition_probability}, closely following the derivation presented in~\cite{Tesser2025}. As our starting point, we consider the unitary evolution from time $0$ to time $t$ of the full Hamiltonian $\hat{H}(t)=\hat{H}_0 + \hat{V}(t)$. The time-evolution operator reads
\begin{equation}
    \hat{\U}(t,0) = \T \exp\left\{-\frac{i}{\hbar}\int_{0}^{t} \hat{H}(s)ds\right\},
\end{equation}
where $\T$ denotes the time ordering. By treating the tunneling Hamiltonian $\hat{V}(t)$ perturbatively, we next expand the full unitary evolution as $\hat{\U}(t, 0) \approx \hat{\U}_0(t,0) + \delta \hat{\U}(t, 0)$, where
\begin{subequations}
\label{eq:timeOP_expanded}
\begin{align}
        \hat{\U}_0(t,0) &\equiv e^{-i\hat{H}_0 t/\hbar}, \\
        \delta \hat{\U}(t,0) &\equiv -\frac{i}{\hbar} \int_{0}^{t} dx \ \hat{\U}_0(t,x) \hat{V}(x) \hat{\U}_0(x,0),
\end{align}
\end{subequations}
are the evolution induced by the free Hamiltonian $\hat{H}_0$ and the first correction due to the tunneling Hamiltonian $\hat{V}(t)$, see Eq.~\eqref{eq:V_Hamiltonian}. For the conditional probability between two common eigenstates $a,b$ of the Hamiltonian $\hat{H}_0$, this means
\begin{eqnarray}
&&    |\braket{b|\mathcal{U}(t,0)|a}|^2  \approx  |\langle b|\hat{\U}_0(t,0) +        \delta \hat{\U}(t,0)|a\rangle|^2\\
&&        \approx \left|\delta_{ab}-i\left[A_{ba}\frac{e^{-i\omega t}-e^{-i\epsilon_{ba}t/\hbar}}{\epsilon_{ba}-\hbar\omega}+A_{ab}^*\frac{e^{i\omega t}-e^{-i\epsilon_{ba}t/\hbar}}{\epsilon_{ba}+\hbar\omega}\right]\right|^2\nonumber
\end{eqnarray}
where we used the matrix elements of the tunneling Hamiltonian $A_{ba}=\langle b|\hat{A}|a\rangle$ and the energy differences $\epsilon_{ba}=\epsilon_b-\epsilon_a$. If we now further assume that the tunneling operator allows for either the transition $|a\rangle\to|b\rangle$ \textit{or} $|b\rangle\to|a\rangle$ but not for both, i.e. $A_{ab}A_{ba}=0$, we find
\begin{equation}
\begin{split}
 |\braket{b|\mathcal{U}(t,0)|a}|^2  &\approx  |A_{ba}|^2\frac{2-2\cos\left[\left(\epsilon_{ba}/\hbar-\omega\right)t\right]}{\left(\epsilon_{ba}-\hbar\omega\right)^2}\\
&+|A_{ab}|^2\frac{2-2\cos\left[\left(\epsilon_{ba}/\hbar+\omega\right)t\right]}{\left(\epsilon_{ba}+\hbar\omega\right)^2}\,,
\end{split}
\end{equation}
for $b\neq a$.
The long-time limit tunneling \textit{rates} from state $a$ to state $b$ are then simply obtained by taking a time derivative, together with the limit $t\to\infty$, while keeping $\frac{|A_{ba}|t}{\hbar},\frac{|A_{ab}|t}{\hbar}\ll1$,
\begin{eqnarray}
 &&\partial_t|\braket{b|\mathcal{U}(t,0)|a}|^2  \xrightarrow{t\to\infty}\nonumber \\ 
 &&\frac{2\pi}{\hbar}|A_{ba}|^2\delta(\epsilon_{ba}-\hbar\omega)
+\frac{2\pi}{\hbar}|A_{ab}|^2\delta(\epsilon_{ba}+\hbar\omega)\ .
\end{eqnarray}
We now choose the transition matrix element $A_{ba}$ to be the non-zero one. The transition rates are obtained by multiplying by the probability of the initial state
\begin{eqnarray}
    \Gamma_{a\to b}  & = & \frac{2\pi}{\hbar}\delta(\epsilon_{ba}-\hbar\omega)|A_{ba}|^2p_a,\label{eq:ab_rate}\\
    \Gamma_{a\gets b} & = & \frac{2\pi}{\hbar}\delta(\epsilon_{ba}-\hbar\omega)|A_{ba}|^2p_b.\label{eq:ba_rate}
\end{eqnarray}
Finally, to get the full transport rates, we multiply by the probability of the initial state and sum over all initial and final states contributing to the process of absorption or emission of $\hbar\omega$, namely $\Gamma_\rightarrow  =  \sum_{a,b} \Gamma_{a\to b}$ and $\Gamma_\leftarrow= \sum_{a,b}\Gamma_{a\gets b}$. Specializing to a bipartite system, with $\hat{H}_0=\hat{H}\L+\hat{H}\R$ and $\hat{H}\L\ket{nl}=\epsilon_{n}^{(\text{L})}\ket{nl}$, $\hat{H}\R\ket{nl}=\epsilon_{l}^{(\text{R})}\ket{nl}$, the rates take the form
\begin{equation}\label{eq:special-rates}
\begin{split}    
    \Gamma_{mk\gets nl} &\equiv \frac{2\pi}{\hbar} \delta(\epsilon_{mn}^\text{(L)}+\epsilon_{kl}^\text{(R)}-\hbar\omega) |A_{mk,nl}|^2  p_l^\text{(R)}p_n^\text{(L)}, \\
    \Gamma_{mk\to nl} &\equiv\frac{2\pi}{\hbar} \delta(\epsilon_{mn}^\text{(L)}+\epsilon_{kl}^\text{(R)}-\hbar\omega) |A_{mk,nl}|^2  p_k^\text{(R)}p_m^\text{(L)},  
\end{split}
\end{equation}
leading to Eq.~\eqref{eq:rate-definition} in the main text.

\section{Current and noise in the weak-tunneling regime}\label{app:current_and_noise}

It is interesting to connect these rates to transport quantities. On one hand, this allows us to exploit our constraints to develop bounds on the noise under a temperature bias in terms of currents. On the other hand, it provides an experimental access to test our bounds on transition probabilities and rates. 

We start by defining a current operator via Heisenberg's equation of motion as
\begin{equation}
    \hat{I}(t) \equiv -\frac{i}{\hbar}\left([\hat{Q},\hat{A}]e^{-i\omega t} + [\hat{Q},\hat{A}^\dagger]e^{i\omega t}\right),\label{eq:general_current-operator}
\end{equation}
where the operator for a generalized charge has the properties $[\hat{H_0}, \hat{Q}]=0$ and $\hat{Q}|a\rangle=q_a|a\rangle$ for the common eigenstates $|a\rangle$. By treating the tunneling Hamiltonian $\hat{V}(t)$ perturbatively, we now study the average current $I = I(t) \equiv \langle \hat{I}_\text{H}(t) \rangle$ and its zero-frequency noise $\S \equiv \int dt\langle \delta \hat{I}_\text{H}(t) \delta \hat{I}_\text{H}(0)\rangle$, where $\hat{I}_\text{H}(t)$ is the current operator in the Heisenberg picture and $\delta \hat{I}_\text{H}(t) \equiv \hat{I}_\text{H}(t) - I(t) $ its fluctuation. 

With Eq.~\eqref{eq:timeOP_expanded}, the average current can be expanded in powers of the tunneling operator $A$ as $\langle \hat{I}_\text{H}(t)\rangle \approx I^{(1)}(t) + I^{(2)}(t)$ with
\begin{subequations}
\label{eq:current_expanded}
\begin{align}
        I^{(1)}(t) &\equiv \Tr{}{\hat{\U}_0^\dagger(t,0)\hat{I}(t)\hat{\U}_0(t,0)\hat{\rho}_0},\label{eq:current_expanded_a}\\
        I^{(2)}(t) &\equiv \Tr{}{[\delta \U^\dagger (t,0)\hat{I}(t)\hat{\U}_0(t,0) + \hat{\U}_0^\dagger(t,0) \hat{I}(t)\delta \U (t,0)]\hat{\rho}_0},\label{eq:current_expanded_b}
\end{align}
\end{subequations}
with $\hat{I}(t)$ given in Eq.~\eqref{eq:general_current-operator} and $\hat{\rho}_0$ is the state of the system at time $t=0$. Connecting to Appendix~\ref{app:weak-transition rates}, we assume that $[\hat{\rho}_0, \hat{H}_0]=0$, such that, if $\{\ket{a}\}$ are the eigenstates of $\hat{H}_0$ with energies $\epsilon_a$, their occupations are given by $\hat{\rho}_0 \ket{a} = p_a\ket{a}$. For the zero-frequency noise, instead we start from the expansion
\begin{equation}
\begin{split}
    \S(t) &\approx \langle \hat{I}_\text{H}(t)\hat{I}(0)\rangle \approx \Tr{}{\hat{\U}_0^\dagger(t,0)\hat{I}(t)\hat{\U}_0(t,0)\hat{I}(0)\hat{\rho}_0}. 
\end{split}
\end{equation}
We now again assume that the tunneling operator satisfies $A_{ab}A_{ba}=0$, which means that only one of the transitions between $\ket{a}\to \ket{b}$ and $\ket{b}\to \ket{a}$ is possible. After some algebra, we find the current and noise expressed in terms of the transition rates~\eqref{eq:ab_rate} and \eqref{eq:ba_rate}
\begin{equation}
    \begin{split}
        I &= \sum_{ab} (q_b-q_a)[\Gamma_{b\to a} - \Gamma_{b\gets a}], \\
        \mathcal{S} &= \sum_{ab} (q_b-q_a)^2[\Gamma_{b\to a} + \Gamma_{b\gets a}], \\
    \end{split}
\end{equation}
where we recall that $q_a\ket{a}=\hat{Q}\ket{a}$.
If we now add the hypothesis $[\hat{Q},\hat{A}]=q\hat{A}$ as for standard charge currents, such that  one has $(q_{b}-q_{a})\to q$, the results for current and noise in terms of rates of  Sec.~\ref{sec:bipartite} in the main text are thus recovered.

\section{Gr\"onwall inequality on out-of-equilibrium rates in a bipartite system}\label{app:Gronwall-inequality}

From the thermodynamic constraint in Eq.~\eqref{eq:thermodynamic-constraint}, it is possible to remove the out-of-equilibrium response of tunneling rates,
$\partial\L\Gamma_{\leftrightarrows}(\beta_\mathrm{h}, \beta_\mathrm{c})$, by means of Gr\"onwall's lemma~\cite{Gronwall1919Jul}.
Indeed, Eq.~\eqref{eq:thermodynamic-constraint} can be cast as
\begin{equation}
   y'(x) \leq f(x) + g(x) y(x),
\end{equation}
where $x$ is the inverse temperature of subsystem L, and
\begin{equation}
    \begin{split}
 y(x) &\equiv \Gamma_\rightleftarrows(x, \beta_\mathrm{c}),\\
 f(x) &\equiv 
 \partial\L\Gamma_{\leftrightarrows}(\beta_\mathrm{c}, \beta_\mathrm{c})
 - \Delta F\L^\text{(c)}(x)\eta^\text{(h)}(x) \Gamma_\rightleftarrows(\beta_\mathrm{c}, \beta_\mathrm{c}),\\ 
 g(x) &\equiv -\Delta F\L^\text{(h)}(x)\eta^\text{(c)}(x).
    \end{split}
\end{equation}
Considering the corresponding homogeneous differential equation,
\begin{equation}
    v'(x) = g(x)v(x)\quad\rightarrow\quad v(x)=\exp\left[\int_x^{\beta_\mathrm{c}} g(s)ds\right],
\end{equation}
we see that $v(\beta_\mathrm{c})=1$ and $v(x)\geq 0$ for all $x$. The derivative of $y(x)/v(x)$ reads
\begin{equation}
    \frac{d}{dx}\frac{y(x)}{v(x)} = \frac{v(x)y'(x)-y(x)v'(x)}{v(x)^2}\leq \frac{f(x)}{v(x)}.
\end{equation}
Integrating this expression from $\beta_\mathrm{h}$ to $\beta_\mathrm{c}$ leads to
\begin{equation}
    y(\beta_\mathrm{c}) -\frac{y(\beta_\mathrm{h})}{v(\beta_\mathrm{h})} \leq \int_{\beta_\mathrm{c}}^{\beta_\mathrm{h}} f(x)\exp\left[-\int_x^{\beta_\mathrm{c}} g(s)ds\right] dx, 
\end{equation}
which is Gr\"onwall's inequality.
Reordering the terms we find
\begin{equation}\label{app:eq:Gronwall-inequality-on-rates}
\begin{split}
    \Gamma_\rightleftarrows(\beta_\mathrm{h},\beta_\mathrm{c})&\geq \Gamma_\rightleftarrows(\beta_\mathrm{c},\beta_\mathrm{c})\exp\left[\int_{\beta_\mathrm{h}}^{\beta_\mathrm{c}} g(x)dx\right] + \\
    &\quad- \int_{\beta_\mathrm{h}}^{\beta_\mathrm{c}} f(x)\exp\left[\int_{\beta_\mathrm{h}}^x g(s)ds\right] dx.
\end{split}
\end{equation}
Notably, the right-hand side of Eq.~\eqref{app:eq:Gronwall-inequality-on-rates}  contains neither the out-of-equilibrium rates nor their derivatives.

\bibliography{refs}

\end{document}